\documentstyle[preprint,aps,eqsecnum]{revtex}

\tighten

\begin{document}
\draft

\title{
Mean-field description of ground-state properties \\
of drip-line nuclei.\\
(I) Shell-correction method
         }
\author  { W. Nazarewicz,$^{a,b,c,d}$,
           T.R. Werner,$^{a,b,c}$ and
           J. Dobaczewski,$^c$}
\address {
           $^a$~Physics Division,  Oak Ridge National Laboratory, \\
                P.O. Box 2008, Oak Ridge, Tennessee 37831, USA\\[1mm]
           $^b$~Department of Physics and Astronomy, University of Tennessee,\\
                Knoxville, Tennessee 37996, USA\\[1mm]
           $^c$~Institute of Theoretical
                Physics, Warsaw University, \\
                ul. Ho\.za 69, 00-681 Warsaw, Poland,\\[1mm]
           $^d$~Joint Institute for Heavy-Ion Research, Oak Ridge,
                Tennessee 37831, USA}

\maketitle

\begin{abstract}
A shell-correction method is
applied
to nuclei far from the beta stability line and its suitability to
describe effects of the particle continuum is discussed.
The sensitivity of predicted
locations of  one- and two-particle drip lines
to details of
the macroscopic-microscopic model  is analyzed.
\end{abstract}


\narrowtext

\section{Introduction}

The theoretical description of drip-line nuclei is one of the most exciting
challenges today. The coupling between bound states and continuum
invites  strong interplay between various aspects of
nuclear structure and reaction theory, and
calculations have
strong astrophysical implications, especially in the context
of  the r-process mechanism \cite{[Kra93]}.

Theoretically, the physics of nuclei with very large values
of neutron or proton excess
is a challenge for well-established models of nuclear structure
and, because of dramatic extrapolations involved,
it invites a
variety of theoretical approaches.
 Since
the parameters of  interactions used in the usual shell-model or mean-field
calculations
are determined so as to reproduce the
properties of beta-stable nuclei, the parameters may not always be proper to
be used in the calculation of drip-line nuclei.
One hopes, however, that spectroscopy of exotic nuclei
by means of a radioactive ion beam technique
will
lead to a better determination of forces, at least
those interaction components that
depend on isospin degrees of freedom.

The closeness of the Fermi level to the particle continuum makes the
theoretical
description of drip-line
nuclei a very challenging task.
To put things in some perspective,
Fig.~\ref{wells} displays the average potential wells,
characteristic of a typical
beta-stable system ($^{120}$Sn), a neutron drip-line system
($^{150}$Sn), and  a proton drip-line system
($^{100}$Sn). While the low-energy structure of $^{120}$Sn is
almost exclusively determined
by the particle-hole or pair excitations
across the Fermi level
from bound states to bound states (corrected for
polarization effects
due to giant vibrations), the lowest particle-hole
or pair modes in drip-line nuclei
are embedded in the particle
continuum. Consequently, any tool of nuclear structure theory
that aims at describing many-body correlations starting from
the mean-field-based
single-particle basis (such as shell-model, BCS, RPA, etc.) has to be modified
in the new regime.

One of the most important nuclear properties is its mass.
The ability of a theoretical model to reproduce
the nuclear binding energy is its ultimate
test; it determines its reliability and practical usefulness.
There exist many  mass calculations
(e.g., those based on the nuclear shell model) focused on a narrow
region of the nuclear chart. These calculations are very successful
in reproducing the data in a given region, but their applicability
to other nuclei is limited. In order to extrapolate far from stability,
the  large-scale global mass calculations are usually used
(see, e.g., reviews \cite{[Mar76],[Hau88]}). Since their
parameters are optimized to reproduce known atomic masses,
it is by no means obvious whether the particle number dependence
predicted by global calculations
at very large (or very small) values of the relative neutron excess
\begin{equation}
 I =\frac{N-Z}{A}
\end{equation}
is correct.
In this context,
a good example is the inclusion of the Coulomb redistribution energy term
in the finite-range droplet model -- strongly motivated by the recent
mass measurements for the heaviest elements \cite{[Mol92a]}.

As far as the placement of nuclear drip lines is concerned,
it is not
the absolute value of nuclear mass but rather
the {\em mass difference} between two isotopes that is of interest.
The difference between binding
energies, $B(Z,N)$, determines both
 the one-neutron separation energy,
$S_n(Z,N) = B_n(Z,N)-B_n(Z,N-1)$, and
the two-neutron separation energy,
$S_{2n}(Z,N) = B_n(Z,N)-B_n(Z,N-2)$. If $S_n$  becomes negative,
then a nucleus is neutron unstable; the condition
$S_n$=0 determines the position
of the one-neutron drip line. By the same token,
if $S_{2n}$ becomes negative, a nucleus is unstable against the emission of
two neutrons; the condition $S_{2n}$=0 determines the position
of the two-neutron drip line. (Proton drip lines are defined
analogously.)
Since $S_n$ and $S_{2n}$ are differences between two masses,
the corresponding statistical  rms deviations
are greater than the estimated rms deviation for the total binding energy.
That is, these quantities are less certain than calculated masses.

Because of their sensitivity to various theoretical details,
predicted drip lines
are strongly model-dependent
\cite{[Mol88a],[Hir91],[Abo92],[Smo93],[Mol93]}. The variations between
predictions
can be attributed to: (i) fundamental differences between
microscopic approaches, (ii) different effective interactions employed
within the same approach, and (iii) different approximations used
when solving the nuclear many-body problem within a given approach.
In this study,
several aspects of nuclear structure at the limits
of extreme isospin are discussed by means of
the macroscopic-microscopic approach.
To test the influence of the particle continuum on
shell corrections and pairing energies,
we use the two versions of the shell-correction method
with the Woods-Saxon average potential
(first version is based on the standard averaging method and
the second version is based on
the semiclassical Wigner-Kirkwood expansion).
In Part II of this work \cite{[Dob94a]},
pairing properties
of drip-line systems and the sensitivity of predictions
to effective
forces are investigated by means of
self-consistent
Hartree-Fock (HF) and  Hartree-Fock-Bogolyubov (HFB)
approaches.

\section{Shell-Correction Method and Particle Continuum}

The main assumption of the
shell-correction (macroscopic-microscopic)  method
\cite{[Swi64],[Str67],[Str68],[Bun72],[Bra72]}
is that the total energy
of a nucleus can be decomposed  into two parts:
\begin{equation}\label{Eshell}
E = \tilde{E}+E_{\rm shell},
\end{equation}
where $\tilde{E}$ is the macroscopic energy (smoothly depending on the
number of
nucleons and thus associated with the ``uniform" distribution of
single-particle orbitals), and $E_{\rm shell}$ is the shell-correction term
fluctuating with particle number reflecting the non-uniformities of
the single-particle level distribution, i.e., shell effects.
In order to make a separation (\ref{Eshell}),
one starts from
the one-body HF  density matrix $\rho$
\begin{equation}\label{rhosp}
{\rho}(x',x) = \sum_{i-occ} \phi_i(x')\phi^*_i(x),~~~x=\left(\bbox{r},
\bbox{\sigma}\right),
\end{equation}
which can be decomposed into a ``smoothed" density $\tilde \rho$
 and a
correction $\delta\rho$, which fluctuates with the shell filling
\begin{equation}\label{rhot}
\rho =\tilde\rho + \delta\rho.
\end{equation}
The density matrix $\tilde\rho$ can be expressed by means of the
smoothed distribution numbers $\tilde{n}_i$
\begin{equation}\label{rhos}
{\tilde\rho}(x',x) = \sum_{i}\tilde{n}_i\phi_i(x')\phi^*_i(x).
\end{equation}
When considered as a function of the single-particle energies $\epsilon_i$,
the numbers $\tilde{n}_i$
vary smoothly in an
energy interval of the order of the energy difference between major shells.
The
$\tilde{n}_i$'s can, in principle, take
values smaller than zero and larger
than unity \cite{[Bun72]}. Consequently, they do not
represent the single-particle  occupations in the strict sense.

In order to justify Eq.~(\ref{Eshell}),
the expectation value of a  HF Hamiltonian
(containing the kinetic energy, $t$, and the two-body
interaction, $\bar v$)
can be written in terms of
$\tilde\rho$ and $\delta\rho$ \cite{[Bun72],[RS80]}:
\begin{equation}\label{O}
E_{\rm HF} = {\rm Tr}(t\rho) + {1\over 2}{\rm Tr\,Tr}
(\rho\bar v\rho) = \tilde E + E_{\rm osc} +
O(\delta\rho^2),
\end{equation}
where
\begin{equation}
\tilde E = {\rm Tr}
(t\tilde\rho) + {1\over 2}{\rm Tr\,Tr}(\tilde\rho\bar v\tilde\rho)
\end{equation}
is the average  part of $E_{\rm HF}$ and
\begin{equation}\label{Eosc}
E_{\rm osc} = {\rm Tr}(\tilde h_{\rm HF}\delta\rho)~~~\left(\tilde
h_{\rm HF}=t+{\rm Tr}(\bar v\tilde\rho)\right)
\end{equation}
is the first-order term in $\delta\rho$ representing the
shell-correction contribution to $E_{\rm HF}$.
If a  deformed average potential
gives a similar spectrum to the averaged HF potential
$\tilde h_{\rm HF}$ then the oscillatory part of $E_{\rm HF}$, given by
 Eq.~(\ref{Eosc}),
is very close to that of the deformed shell model,
$E_{\rm shell}$$\approx$$E_{\rm osc}$.
 The
second-order term
in Eq.~(\ref{O}) is usually very small and can be neglected \cite{[Bra81]}.
The above relation, known as the {\em Strutinsky energy theorem},
makes it possible to calculate the total energy using the
non-self-consistent, deformed
independent-particle
model;
the average part $\tilde E$ is usually replaced by the corresponding
phenomenological liquid-drop
(or droplet) model value, $E_{\rm macr}$.
It is important that $E_{\rm shell}$ must not contain any regular
(smooth) terms analogous
to those already included in the phenomenological macroscopic part.
(The extension of the energy theorem
to the case with pairing is straightforward. The resulting expression
for shell correction  contains an
additional  contribution from the fluctuating
part of the pairing energy, see Sec.~\ref{Sec_pairr}.)

There are  two
 single-particle level densities that define the shell correction.
The   (deformed) shell-model single-particle level
density
\begin{equation}\label{gg}
g(\epsilon) = \sum_{i}\delta(\epsilon-\epsilon_i)
\end{equation}
gives the single-particle energy, $E_{\rm s.p.}$.
The smooth single-particle energy, $\tilde{E}_{\rm s.p.}$, is given  by
the mean single-particle level density,
$\tilde{g}(\epsilon)$,
obtained
from $g(\epsilon)$
by folding with a smoothing function
$f(x)$:
\begin{equation}\label{ggsmooth}
\tilde{g}(\epsilon) = {1\over\gamma}
\int_{-\infty}^{+\infty}d\epsilon'g(\epsilon')
f\left(\frac{\epsilon-\epsilon'}{\gamma}\right)=
{1\over\gamma}\sum_{i}f\left(\frac{\epsilon-\epsilon_i}{\gamma}\right).
\end{equation}
In Eq.~(\ref{ggsmooth}),  $\gamma$ is the smoothing range; it should be
larger than the typical distance between major shells.
As follows from Eqs.\,(\ref{Eosc})--(\ref{ggsmooth}),
the shell correction  can be calculated by taking the difference
between the sum of occupied levels and its average value, i.e.
\begin{equation}\label{shell}
E_{\rm shell} = E_{\rm s.p.} - \tilde{E}_{\rm s.p.} =
\sum_{i-{\rm occ}}\epsilon_i - \int_{-\infty}^{\tilde{\lambda}}
\epsilon\,\tilde{g}(\epsilon)d\epsilon,
\end{equation}
where $\tilde{\lambda}$ is the smoothed Fermi level defined through
the particle number equation:
\begin{equation}\label{ltilde}
N = \int_{-\infty}^{\tilde{\lambda}} \tilde{g}(\epsilon)d\epsilon.
\end{equation}

The folding function $f(x)$ can be written as a product
\begin{equation}
f(x)=\omega(x)P_p(x),
\end{equation}
where $\omega(x)$ is a weighting function and $P_p(x)$
is the so-called curvature-correction polynomial of the $p$th order.
The smoothing procedure should be {\em unambiguous}, i.e.,
the averaging should extract only the fluctuating part, leaving
the smooth part untouched. This condition defines $P_p(x)$ for
any specific choice of $\omega(x)$ . For instance, for
the infinite potential well and a Gaussian weighting function,
$\omega(x)=\pi^{-1/2}{\exp}(-x^2)$, the curvature-correction
polynomial can be be expanded in a finite
series of Hermite polynomials of
even order \cite{[Bun72],[Bol72]}
\begin{equation}\label{curvature}
P_p(x) = \sum_{k=0,2,...}^{p}\frac{(-1)^{k/2}}{2^k(k/2)!}H_k(x).
\end{equation}
(For other examples of $f(x)$, see Ref.~\cite{[Bra73]}.)

The smoothed single-particle energy can be expressed in the
form \cite{[Bra73]}:
\begin{equation}\label{Eplateau}
\tilde{E}_{\rm s.p.} =
\int_{-\infty}^{\tilde\lambda}
\epsilon\, \tilde{g}(\epsilon) d\epsilon=
\sum_{i}\epsilon_i \tilde{n}_i + \gamma \frac{d\tilde{E}_{\rm s.p.}}{d\gamma},
\end{equation}
where the smoothed distribution numbers are
\begin{equation}
\tilde{n}_i =
{1\over\gamma}\int_{-\infty}^{\tilde{\lambda}}d\epsilon\,
f\left(\frac{\epsilon-\epsilon_i}{\gamma}\right).
\end{equation}
Since the value of $\tilde{E}_{\rm s.p.}$
should not depend on the smoothing range $\gamma$
(nor on the order of curvature correction
$p$), the second term in Eq.~(\ref{Eplateau}) must vanish, i.e.
\begin{equation}\label{plateau}
\frac{d\tilde{E}_{\rm s.p.}}{d\gamma}=0
{}~~~~\left(\frac{d\tilde{E}_{\rm s.p.}}{dp}=0\right).
\end{equation}
If the above
{\em plateau condition} does not hold, the Strutinsky averaging method
does not yield an unambiguous result.

The treatment of the particle continuum in the shell-correction approach is
an old problem.
For finite-depth potentials,
the sum in Eq.~(\ref{ggsmooth}) should be replaced
by a sum over bound single-particle states
and an integral over positive-energy single-particle continuum, i.e.
\begin{equation}\label{sume}
\sum_i  \rightarrow
 \sum_{i-bound} + \int_{0}^{\infty}d\epsilon\,\Delta g_{\rm cont}(\epsilon),
\end{equation}
where $\Delta g_{\rm cont}$ is the continuum shell level density
 \cite{[Bet37],[Pet71]}:
\begin{equation}
\Delta g_{\rm cont}(\epsilon) =
{1\over \pi}\frac{d\delta_{\epsilon}}{d\epsilon},
\end{equation}
and $\delta_{\epsilon}$ is the corresponding phase shift (calculated
for all partial waves).

Already in 1970 Lin \cite{[Lin70]} pointed out
that the contribution from the particle
continuum can affect the value of $E_{\rm shell}$, even
for nuclei at the beta-stability line. In particular,
he considered the positive-energy continuum in the energy interval
$0<\epsilon<20$\,MeV to calculate
the neutron shell correction
for $^{208}$Pb.
No plateau in $\tilde{E}_{\rm s.p.}$
was obtained and the result turned out be strongly
$\gamma$- and $p$-dependent. Soon afterwards
Ross and Bhaduri \cite{[Ros72]} carried out calculations
 along the same lines as in Ref.~\cite{[Lin70]}
and demonstrated that,
 by taking into account contributions from
all neutron resonances up to $\sim$100 MeV in $^{208}$Pb, the
 plateau condition (\ref{plateau}) could be  met.

Bolsterli {\em et al.} \cite{[Bol72]} made an attempt
to simulate the effect
of the continuum by using the {\em quasibound states}, i.e., the states
resulting from the diagonalization of a finite potential in a large
harmonic oscillator basis. The authors suggested
 a ``working" prescription
(the number of oscillator shells should not be too large,
$N_{\rm osc}$$\sim$12), but no systematic analysis was carried out.
(Later it was shown in Ref.~\cite{[Sob77]} that
the inclusion of quasibound states
requires relatively high correction orders.) In the following,
the Strutinsky averaging
procedure including quasibound states will be referred to as
the standard averaging method (SAM).

There have been some suggestions on how to generalize the Strutinsky
averaging procedure for finite potentials. Bunatian {\em et al.}
\cite{[Bun72]}
exploited only the bound states of a
finite-depth single-particle potential
and derived modified expressions for the curvature correction. This
method was then improved and exploited in series of papers
by Strutinsky and Ivanyuk
\cite{[Str75],[Iva78],[Iva79]}.
For instance, Ref.~\cite{[Iva79]} demonstrates
 that the standard way of computing shell corrections,
based on quasibound states, leads to
serious errors in theoretical predictions for
masses, fission barriers, and
deformation energies. They also pointed out that the
uncontrolled smooth component
in shell
corrections, resulting from the incorrect treatment of continuum, can
cause some
renormalization of the parameters of the macroscopic energy formula,
determined from fitting the nuclear masses. Unfortunately, the
results discussed in Ref.~\cite{[Iva79]} suggest that the
renormalization procedure, based
solely on bound states,
produces very unstable shell corrections in cases
when the upper limit of the averaging interval
(i.e., the number of the highest single-particle level considered)
approaches the
actual number of particles. Actually, such a situation happens at the
particle drip lines.

The proper treatment of resonances is
not an easy task, especially for
deformed systems. Therefore,
other methods of dealing with continuum are more useful
  in practical applications.
In 1973 Jennings
 proved \cite{[Jen73]}  the equivalence between the Strutinsky
 approach and
the semiclassical averaging based on the partition function
 method (Wigner-Kirkwood expansion).
In the semiclassical
approximation, the Fermi energy $\lambda_{\rm sc}$ is determined by
\begin{equation}\label{Npart}
N = {\cal L}^{-1}_{\lambda_{\rm sc}}\left(\frac{Z(\beta)}{\beta}
\right),
\end{equation}
and  the smoothed energy of the system is
\begin{equation}\label{part}
\tilde{E}_{\rm sc} = N\lambda_{\rm sc}
 - {\cal L}^{-1}_{\lambda_{\rm sc}}\left(\frac{Z(\beta)}{\beta^2}
\right),
\end{equation}
where ${\cal L}$ denotes the Laplace transform and $Z$ is
the semiclassical partition function \cite{[Uhl36]}
\begin{equation}
Z(\beta) = \frac{2}{h^3}\int {\rm e}^{-\beta H_c}
\left(1 + \hbar w_1 + \hbar^2 w_2 + ...\right) d^3pd^3r,
\end{equation}
where $H_c$ is the classical one-body Hamiltonian of the system
and the $w$'s are defined in terms of the one-body potential.
The partition function method was shown to be an excellent tool
for computing the average single-particle energy
\cite{[Jen75],[Jen75a],[Jen75b],[Bra76]}. In particular,
Ref.~\cite{[Jen75]} contains the  explicit numerical check of the
equivalence between
the Wigner-Kirkwood (WK) expansion and
the full Strutinsky averaging with the correct treatment of resonances.
(For more discussion, see Ref.~\cite{[Jen75a]}, Table I.)

\section{Woods-Saxon model and the particle continuum}

The deformed shell model used in this study
is assumed to be of the form of the average deformed Woods-Saxon
(WS) potential, which contains a central part, a spin-orbit term,
and a
Coulomb potential for protons. All these terms depend explicitly on a set of
external (axial)
deformation parameters, $\beta_\lambda$, defining the nuclear surface:
\begin{equation}\label{Ro}
R(\theta; \bbox{\beta}) = C(\bbox{\beta}) r_\circ A^{1/3} \left[ 1 +
\sum_{\lambda} \beta_{\lambda} Y_{\lambda 0} (\theta)
\right],
\end{equation}
where the coefficient $C$ assures that the total volume enclosed by
the surface
(Eq.~(\ref{Ro})) is conserved.

The deformed WS  potential is assumed in the form \cite{[Cwi87]}:
\begin{mathletters}
\label{WS}
\begin{equation}
V(\bbox{r}) = V_\circ\left[1+\kappa_\circ I\right]f(\bbox{r}),
\end{equation}
\begin{equation}
f(\bbox{r})=\frac{1}{1+{\exp}\left[\xi(\bbox{r})/a\right]},\label{formfactor}
\end{equation}\end{mathletters}\noindent%
where the function $\xi(\bbox{r})$ is equal to the
(perpendicular) distance (taken with the minus
sign inside the surface) between the point $\bbox{r}$ and the nuclear
surface represented by Eq.~(\ref{Ro}). The diffuseness parameter
$a$ is independent of $Z$ and $N$. This choice of the WS potential
guarantees that the diffuse region of the potential is constant independently
of the nuclear deformation.

The deformed spin-orbit (SO) potential is
\begin{equation}\label{SO}
V_{\rm SO}(\bbox{r}) = -\frac{\kappa}{M}\left(\bbox{\nabla}f_{\rm SO}
 \times \bbox{p}\right)
\cdot \bbox{s}.
\end{equation}
Here $\kappa$ is a (dimensionless) SO strength factor and
\begin{equation}\label{distSO}
f_{\rm SO}(\bbox{r}) = \frac{1}
              {1+{\exp}\left[\xi_{\rm SO}(\bbox{r})/a_{\rm SO}\right]}
\end{equation}
is a WS form factor. (Since the radius of the SO potential,
$r_{\circ-{\rm SO}}$,
is, in general, different from the radius of the central part,
the distance function
entering Eq.~(\ref{distSO}) is indicated by  $\xi_{\rm SO}$.)

The Coulomb potential for
protons, $V_{\text{C}}$, is assumed to be that of the charge
$(Z-1)e$ distributed with
 the charge density $\rho_{\text{C}}(\bbox{r})=\rho_\circ f(\bbox{r})$,
where $f(\bbox{r})$
is the WS form factor of Eq.~(\ref{formfactor}).
In our study we employed the set of WS
parameters introduced in Ref.~\cite{[Dud81]}. These parameters
have been widely used in nuclear structure calculations around the
beta-stability line and for neutron-rich nuclei. So far,
no attempt has been made
to optimize these parameters for very neutron-rich nuclei.

The eigenstates of the WS Hamiltonian
\begin{equation}\label{WSH}
H_{\text{WS}} = T + V + V_{\rm SO} + \case{1}{2}\left(1+\tau_3\right)
                V_{\text{C}}
\end{equation}
are found by means of the expansion method in the
axially deformed harmonic-oscillator basis. The oscillator
frequencies, $\hbar\omega_\perp$ and $\hbar\omega_z$,
have been optimized according to Ref.~\cite{[Cwi87]}.
In our calculations we used {\em all}
basis states belonging to $N$$\leq$$N_{\rm osc}$ (stretched) oscillator shells.

The pattern of eigenstates
of the WS Hamiltonian
(\protect\ref{WSH}),
 as a function of the number of
harmonic-oscillator quanta included in the basis,
(10$\leq$$N_{\rm osc}$$\leq$50)
is displayed in Fig.~\ref{Fig_Sn}.
The left portion
shows the results representative of the spherical shape.
[Note that every single-particle orbital is $(2j+1)$-fold
degenerate.] The spherical symmetry is lifted
in the right portion, representative of the deformed situation
($\beta_2$=0.4; here each orbital
is only two-fold degenerate due to the time reversal symmetry).
It is seen that the energies of bound states ($\epsilon_i<0$)
converge very rapidly with the size of the basis; the convergence
 is achieved
for $N_{\rm osc}$$\approx$14. The positive-energy quasibound states
representing the discretized
particle continuum vary dramatically with $N_{\rm osc}$.
Asymptotically, as the basis becomes infinite, quasibound  states approach
zero energy (cf. Fig.~18 of Ref.~\cite{[Bol72]} and related discussion).
This leads to an
increased single-particle level density at $\epsilon_i>0$
at large $N_{\rm osc}$-values. The only states which are not
strongly affected in the considered interval
of $N_{\rm osc}$ are the high-$j$ orbitals
which are fairly well localized inside
a pocket in the centrifugal barrier created at positive energies
by the mean-field potential.

The situation in protons is slightly less
dramatic  than in neutrons. Because of the presence of
the Coulomb barrier ($\sim$9 MeV in $^{120}$Sn), the positive-energy
single-particle states below $\sim$5\,MeV are fairly stable. (They
represent  narrow sub-barrier resonances, interesting in the context
of proton emitters.)
Judging from the
results shown in Fig.~\ref{Fig_Sn}, it is impossible to
find an ``optimal" value of $N_{\rm osc}$, at  which the positive-energy
quasibound spectrum would become  a fair representation
of  physical continuum (resonances).

Since, at the particle drip lines, the radial asymptotics of
wave functions
are particularly important,
it is instructive  to relate $N_{\rm osc}$ to the radial
dimensions of the harmonic oscillator wave function.
Its size
depends both on the principal quantum number, $N$, and on the orbital
quantum number, $\ell$. The classical major axis of the orbit ($N, \ell$)
is given by \cite{[Boh69]}
\begin{equation}\label{orbit}
(r_{\rm max})^2 \approx L^2
                \left[N+\sqrt{N^2-\ell^2}\right],
\end{equation}
where $L$=$\sqrt{\frac{\hbar}{M\omega_\circ}}$ is the oscillator length,
i.e.,  for the $\ell$=0 oscillator states,
 the classical outer turning point
is
$r_{\rm max}$$\approx$$L\sqrt{2N}$.
A weak dependence of $r_{\rm max}$ on $N$ makes it very difficult,
if not impossible, to  describe the asymptotic
radial behavior of the wave function
using the expansion method in the harmonic oscillator basis.
{}From this point of view, the direct integration of the Schr\"odinger
equation in a spatial
box is superior. For instance, for the nucleus $^{120}$Sn
discussed in Fig.~\ref{Fig_Sn}, the oscillator length $L$ appearing in
Eq.~(\ref{orbit}) equals to about 2\,fm, hence
$r_{\rm max}$$\simeq$10\,fm for $N$=12
and $r_{\rm max}$$\simeq$20\,fm for $N$=50.
(For a pedagogical example, see Fig.~19 of Ref.~\cite{[Bol72]}.)
By the same token, for a high-$\ell$ orbit with $\ell$=$N$
 the major axes are
$r_{\rm max}$$\simeq$7\,fm and 14\,fm for $N$=12 and $N$=50, respectively.

\section{Wigner-Kirkwood semiclassical expansion}

The semiclassical approximation for a one-body Hamiltonian,
including SO interaction, was developed in Ref.~\cite{[Jen75]}.
Denoting by $U(\bbox{r})$ a one-body potential without the SO
interaction,
\begin{equation}
U(\bbox{r}) = V(\bbox{r}) + \case{1}{2}\left(1+\tau_3\right)
              V_{\text{C}}(\bbox{r}),
\end{equation}
the particle number equation (\ref{Npart}) takes the form:
\widetext
\ifpreprintsty
\begin{eqnarray}\label{Npart1}
N = \frac{1}{3\pi^2}\left(\frac{2M}{\hbar^2}\right)^{3\over 2}
\int^{\bbox{r}_{\rm sc}} d^3r\biggl\{(\lambda_{\rm sc}-U)^{3\over 2} & + &
\frac{\hbar^2}{2M}\left[\case{3}{4}\kappa^2(\bbox{\nabla}f_{\rm SO})^2
(\lambda_{\rm sc}-U)^{1\over 2}\right.\nonumber\\
 & - & \left.\case{1}{16}\nabla^2
U(\lambda_{\rm sc}-U)^{-{1\over 2}}\right]\biggr\}.
\end{eqnarray}
\else
\begin{eqnarray}\label{Npart1}
N = \frac{1}{3\pi^2}\left(\frac{2M}{\hbar^2}\right)^{3\over 2}
\int^{\bbox{r}_{\rm sc}} d^3r\biggl\{(\lambda_{\rm sc}-U)^{3\over 2} & + &
\frac{\hbar^2}{2M}\left[\case{3}{4}\kappa^2(\bbox{\nabla}f_{\rm SO})^2
(\lambda_{\rm sc}-U)^{1\over 2}\right.
   -   \left.\case{1}{16}\nabla^2
U(\lambda_{\rm sc}-U)^{-{1\over 2}}\right]\biggr\}.
\end{eqnarray}
\fi
The integral is cut off at the
 classical turning point (more precisely: the
surface of turning points)
defined by the relation
$U(\bbox{r}_{\rm sc})=\lambda_{\rm sc}$. The equation (\ref{Npart1})
for $\lambda_{\rm sc}$ is solved iteratively.
The expression (\ref{part}) for the smoothed energy is slightly more
complicated. In analogy to the notation introduced in Ref.~\cite{[Jen75]}, we
may write:
\begin{equation}\label{part1}
\tilde{E}_{\rm sc}  =
\lambda_{\rm sc}N - \left[E^0_{-3} + E^0_{-1} + E^0_{+1}\right] -
\left[E^{\rm SO}_{-1} + E^{\rm SO}_{+1}\right],
\end{equation}
where
\begin{eqnarray}
E^0_{-3} & = & \frac{2}{15\pi^2}\left(\frac{2M}{\hbar^2}\right)^{\frac{3}{2}}
\int^{\bbox{r}_{\rm sc}} d^3r(\lambda_{\rm sc}-U)^{5\over 2},\\
E^0_{-1} & = & - \frac{1}{24\pi^2}\left(\frac{2M}{\hbar^2}\right)^{1\over 2}
\int^{\bbox{r}_{\rm sc}} d^3r(\lambda_{\rm sc}-U)^{1\over 2}\nabla^2U,\\
E^0_{+1} & = & \frac{1}{5760\pi^2}\left(\frac{\hbar^2}{2M}\right)^{1\over 2}
\int^{\bbox{r}_{\rm sc}} d^3r\frac{1}{(\lambda_{\rm sc}-U)^{1\over 2}
(\bbox{\nabla}U)^2}\left\{-7\nabla^4U(\bbox{\nabla}U)^2 + 5(\nabla^2U)^3
\right.\nonumber\\
 & + & 10[\bbox{\nabla}U\cdot\bbox{\nabla}(\nabla^2U)]\nabla^2U
-5(\nabla^2U)^2\bbox{\nabla}U\cdot\bbox{\nabla}(\bbox{\nabla}U)^2
/(\bbox{\nabla}U)^2\nonumber\\
 &+& \left.
(\nabla^2U)\nabla^2(\bbox{\nabla}U)^2 +
\bbox{\nabla}U\cdot\bbox{\nabla}\nabla^2(\bbox{\nabla}U)^2 -
\nabla^2(\bbox{\nabla}U)^2\bbox{\nabla}U\cdot\bbox{\nabla}
(\bbox{\nabla}U)^2/(\bbox{\nabla}U)^2
\right\},\\
E^{\rm SO}_{-1} & = & \frac{\kappa^2}{6\pi^2}\left(\frac{2M}
{\hbar^2}\right)^{1\over 2}
\int^{\bbox{r}_{\rm sc}} d^3r(\lambda_{\rm sc}-U)^{3\over 2}
(\bbox{\nabla}f_{\rm SO})^2,\\
{\rm and}  & & \nonumber \\
E^{\rm SO}_{+1} & = & \frac{\kappa^2}{48\pi^2}
\left(\frac{\hbar^2}{2M}\right)^{\frac{1}{2}}
\int^{\bbox{r}_{\rm sc}} d^3r(\lambda_{\rm sc}-U)^{1\over 2}\biggl\{
\case{1}{2}\nabla^2(\bbox{\nabla}f_{\rm SO})^2-
(\nabla^2f_{\rm SO})^2\nonumber\\
&  +  & \bbox{\nabla}f_{\rm SO}\cdot\bbox{\nabla}(\nabla^2f_{\rm SO})
 -  \frac{(\bbox{\nabla}f_{\rm SO})^2\nabla^2U}{2(\lambda_{\rm sc}-U)}
-2\kappa \Bigl[(\bbox{\nabla}f_{\rm SO})^2\nabla^2f_{\rm SO}\nonumber\\
& - &
\case{1}{2}\bbox{\nabla}f_{\rm SO}
\cdot\bbox{\nabla}(\bbox{\nabla}f_{\rm SO})^2
\Bigr]  +  2\kappa^2(\bbox{\nabla}f_{\rm SO})^4\biggr\}.
\end{eqnarray}
Some technical details of the calculations of the above integrals,
specific to our WS model, are discussed in Appendices A and B.
\narrowtext

The WK expansion ((\ref{Npart1})-(\ref{part1})) is
only applicable
for one-body potentials whose first and higher derivatives exist.
This means that the method cannot be directly applied
to a wide class of single-particle models based on the average
field obtained by means of folding over a sharp generating potential
or a sharp density distribution. Examples here are the
folded Yukawa potential \cite{[Bol72]} or the Coulomb potential
generated by the uniform density distribution; both are continuous
in their values and first derivatives, but are discontinuous
in their second and higher derivatives.

\section{Results}

\subsection{Quasibound states in the shell-correction method}

The dependence of the shell correction
(\ref{shell}) on the size
of the oscillator basis and deformation is shown in Fig.~\ref{Fig_ZrN}
for a neutron-rich system ($^{100}$Zr) and a neutron drip-line nucleus
($^{122}$Zr). In the calculations all the quasibound states up to
a cut-off energy of
131.2\,MeV/$A^{1/3}$ above the Fermi level were
included. As expected, due to the effect of increased density of
quasibound states with $N_{\rm osc}$, the shell correction is
unstable.
It decreases or increases linearly with $N_{\rm osc}$, depending
on the position of the Fermi level.
Indeed, for $^{100}$Zr the value of $E_{\rm shell}$ slowly increases with
$N_{\rm osc}$, while the opposite effect is seen for $^{122}$Zr.
The reason for this different behavior is explained in Fig.~\ref{gavr}
which shows the mean single-particle level density
(\ref{ggsmooth}) as a function of single-particle energy
$\epsilon$ for $^{100,122}$Zr. For relatively low values of
$N_{\rm osc}$ (e.g., $N_{\rm osc}$=12--14), the average level density
increases monotonically and it depends weakly on $p$ (results
for $p$=6 and $p$=12 are very similar). On the other hand, for
$N_{\rm osc}$=50,  there appears an artificial  local minimum in
$\tilde{g}(\epsilon)$ located
at --7$\lesssim$$\epsilon$$\lesssim$--5\,MeV which is
caused by a rapid change in slope of
$\tilde{g}(\epsilon)$ due to
a large number of quasibound states at $\epsilon$$\gtrsim$0.
This fluctuation is caused by a polynomial correction;
as seen in Fig.~\ref{gavr} the results for
$N_{\rm osc}$=50 strongly depend on $p$.
The smooth single-particle energy is obtained by
integrating the product
$\tilde{g}(\epsilon)\epsilon$ up to the Fermi level $\tilde\lambda$, cf.
Eq.~(\ref{Eplateau}). Since
$\tilde{g}(\tilde\lambda)_{N_{\rm osc}=14} <
\tilde{g}(\tilde\lambda)_{N_{\rm osc}=50}$
for $^{122}$Zr (opposite holds for
$^{100}$Zr), this explains the tendency seen in
Fig.~\ref{Fig_ZrN}.

The instability of the shell correction with respect to $N_{\rm osc}$
means that the total mass of a drip-line nucleus cannot be estimated
by means of the SAM.
Moreover, since in different isotopes one obtains
a different dependence on  $N_{\rm osc}$, the two-neutron separation
energies are also affected. In fact,
as seen in Fig.~\ref{Fig_ZrN} the
deformation energies can also be contaminated by the $N_{\rm osc}$-dependence.
The relative values of $E_{\rm shell}$, computed
at different deformations, vary with $N_{\rm osc}$
up to $N_{\rm osc}$=14 ($^{100}$Zr) and
$N_{\rm osc}$=18 ($^{100}$Zr), while
the recommended value of $N_{\rm osc}$ is
12-14 \cite{[Bol72]}.

The results presented in this section demonstrate that
the uncontrolled error in $E_{\rm shell}$ can
seriously affect theoretical
mass predictions for nuclei far from stability.
If parameters of the macroscopic energy  formula are determined by the
global fit to experimental nuclear masses,
some part
of the unwanted effect of the
continuum is absorbed by the isospin-dependent
terms of the liquid-drop (or droplet) model mass formula.

\subsection{Plateau condition in the shell-correction method}

In Fig.~\ref{plateaup}, the total spherical shell corrections
for $^{100}$Zr and $^{122}$Zr
(i.e., including proton
and neutron contributions) are plotted as a function of $\gamma$
for different values of $p$.
For the basis size
we took the recommended value of $N_{\rm osc}$=12.
Neither for $^{100}$Zr nor for
$^{122}$Zr
is the
plateau condition (\ref{plateau})
fulfilled exactly.
 However,
while in the former case the fluctuations of $E_{\rm shell}$
with $\gamma$ and $p$ are rather weak (for the ``usual" values of
$p$=6,\,8 and $\gamma$/$\hbar\omega_\circ$=1.0--1.2
the local plateau in $E_{\rm shell}$
is at $\sim$6.7\,MeV), in the
latter case the variations in shell correction are more
dramatic and no unambiguous values  can be extracted.

In the recent state-of-the-art large-scale mass calculations based on the
folded Yukawa potential, M\"oller {\it et al.}
\cite{[Mol93]} used $\gamma$=$\hbar\omega_\circ B_s$ ($B_s$ is the
ratio of the actual nuclear surface  to the spherical surface, i.e.,
$B_s$=1 at spherical shape),
$p$=8, and $N_{\rm osc}$=12. They noted that, in contrast to
light nuclei which have
very low density of single-particle
levels,
the plateau condition is
fulfilled for  heavy nuclei. Our analysis suggests that
for nuclei far from stability
the plateau
condition can also be violated.

\subsection{Shell energies for the Wigner-Kirkwood smooth energy}

In order to estimate the continuum-related uncertainty in
 the shell energy
in typical SAM calculations,
we performed a  semiclassical analysis
using the WK expansion. Figure~\ref{WKS}
shows  neutron
shell corrections $E_{\rm shell}$(SAM)=$E_{s.p}$$-$$\tilde{E}_{s.p}$
computed by means
of the SAM, Eq.~(\ref{Eplateau}),
($N_{\rm osc}$=12, $\gamma$=$\hbar\omega_\circ$, $p$=8)
as compared to $E_{\rm shell}$(WK)=$E_{s.p}$$-$$\tilde{E}_{\rm sc}$ where
the smoothed WK energy (\ref{part1}) is used.
In both cases the same
WS model was used.
The calculations were carried out for the well-bound neutron systems
$^{100}$Zr and $^{166,186}$Os ($\lambda_n$$<$--6\,MeV),
and for neutron drip-line nuclei
$^{120,124}$Zr and $^{206,226}$Os (--4\,$<$$\lambda_n$$<$--0.2\,MeV).
The average deviation between
${E}_{\rm shell}$(SAM) and ${E}_{\rm shell}$(WK)
ranges
from $\sim$0.4\,MeV in $^{100}$Zr and $^{166}$Os
to $\sim$4.6\,WeV in $^{124}$Zr and $\sim$7.6\,MeV in $^{226}$Os.
Generally, for the neutron-rich systems,
the shell corrections obtained by means of the SAM are
lower than the semiclassical ones and the difference increases with $N$.
Therefore,
in the standard
shell-correction calculations the nuclei around the neutron
drip line are overbound. Another interesting result
presented in Fig.~\ref{WKS} is that
the difference between
${E}_{\rm shell}$(SAM) and ${E}_{\rm shell}$(WK)
depends rather weakly on deformation. This suggests that
 one can approximate $E_{\rm shell}$ by
\begin{equation}\label{renorm}
E_{\rm shell}(\bbox{\beta}) = E_{\rm shell}(\bbox{\beta};{\rm SAM})
+ E_{\rm shell}(\bbox{\beta}=0;{\rm WK})
- E_{\rm shell}(\bbox{\beta}=0;{\rm SAM}),
\end{equation}
where the spherical semiclassical shell correction,
$E_{\rm shell}(\bbox{\beta}=0;{\rm WK})$,
involves calculating one-dimensional radial integrals only.
The renormalization (\ref{renorm}) can be particularly useful in the
large-scale global mass calculations, such as those of
Ref.~\cite{[Mol93]}.
However, one should keep in mind that this renormalization will
not cure the problems related to the plateau condition and the dependence
on $N_{\rm osc}$ discussed above.

It is worth noting that in the
Extended-Thomas-Fermi-with-Strutinsky-integral approach
\cite{[Abo92],[Chu77],[Dut86],[Pea91]} the smooth
energy entering their shell correction is defined by the
semiclassical method \cite{[Dut86]}.
Consequently, their shell corrections are free from the
problems of the continuum.

\subsection{Location of particle drip-lines in the
shell-correction method}\label{Sec_pairr}

In the macroscopic-microscopic method, the second contribution to
the shell energy comes from pairing.
Pairing correlations play a very special role in drip-line nuclei.
This is seen from approximate HFB relations between the Fermi level,
$\lambda$, pairing gap, $\Delta$, and the particle
separation energies. For instance, the neutron separation energies
$S_n$ and $S_{2n}$ are given by \cite{[Smo93],[Bei75a]}
\begin{eqnarray}
S_n & \approx & -\lambda_n - \Delta,\label{separ1n}\\
S_{2n}  & \approx & -2\lambda_n,\label{separ2n}
\end{eqnarray}
which leads to the following conditions for the one-neutron and
two-neutron drip-lines:
\begin{eqnarray}
S_n=0 & \Longrightarrow & \lambda_n + \Delta =0,  \label{drip1n}\\
S_{2n}=0   & \Longrightarrow & \lambda_n=0 \label{drip2n}.
\end{eqnarray}
In particular, the condition (\ref{drip1n}) nicely illustrates
the crucial role of pairing interaction for determining one-neutron
drip line; it shows the equal importance of the single-particle
field characterized by $\lambda$ (determined by the particle-hole
component of the effective interaction) and the pairing field, $\Delta$
(determined by the particle-particle part of the effective interaction).
In fact, just around the particle drip lines,  particle-hole and
particle-particle channels are very strongly coupled, and the standard
procedure based on a ``two-step" treatment of the pairing Hamiltonian
(i.e.,
computing pairing correlations {\em after} determining the
single-particle basis) seems inappropriate.

In part II of our study \cite{[Dob94a]},
the pairing properties of drip-line nuclei are discussed using
the self-consistent theory. Here, we only concentrate on the
standard BCS or Lipkin-Nogami (LN) treatment of pairing correlations,
which is used
in the currently available
large-scale mass calculations \cite{[Abo92],[Mol93]}.

The macroscopic part of the total
energy, $E_{\rm macr}$,
 already contains the average pairing energy, which accounts for the main
part of the even-odd mass difference. Therefore, it is the fluctuating part
of the pairing energy
\begin{equation}\label{pener}
E_{\rm shell}^{\rm (pair)}=E_{\rm pair}-\tilde{E}_{\rm pair}
\end{equation}
that enters the expression for the total shell energy. The pairing
correlation energy, $E_{\rm pair}$, is usually computed
using a
monopole pairing interaction with a constant
(state-independent) strength $G$.

In this study,  the BCS and LN
  equations  were solved considering   the $N_p$ (=$Z$ or $N$)
 lowest single-particle
  orbitals.
The average pairing strength, $G_{\rm avg}$, and the
 average pairing energy,
$\tilde{E}_{\rm pair}$, were
calculated according to the average gap method
\cite{[Bra72],[Bol72],[Bel59]}
in the version of Ref.~\cite{[Mol92b]}.
We have also performed calculations using the fitted pairing strength
$G_{\rm fit}$:
\begin{equation}\label{Gfit}
G_{\rm fit}\cdot A = G_\circ + G_1\,I,
\end{equation}
where the constants $G_\circ$ and $G_1$ are obtained by a fit
to odd-even mass differences in a given region of nuclei.

There is an obvious
 advantage of using $G_{\rm avg}$ rather than $G_{\rm fit}$ when
going far from stability. Since $G_{\rm avg}$ is inversely proportional
to the average single-particle level density at the Fermi level,
$G_{\rm avg} \propto \tilde{g}(\tilde{\lambda})^{-1}$,
this attenuates the influence of quasibound states on
pairing properties. Figure~\ref{DELTAG} shows the
neutron pairing gaps for the tin isotopes computed within the BCS
approximation using $G_{\rm avg}$ or
$G_{\rm fit}$. For neutron numbers
50$\leq$$N$$\lesssim$84 both prescriptions for $G$
yield fairly similar results. However, for higher neutron numbers
the pairing gap obtained with
$G_{\rm fit}$ is much larger that that computed with
$G_{\rm avg}$,
 and the result strongly depends on the size of the single-particle
basis.
In particular, for $N_{\rm osc}$=40
the density of quasibound states is so large that $\Delta_n$$\ne$0 for
$N$=82.
Since the pairing
correction (\ref{pener}) behaves as $\Delta^2$, the effect on
$E_{\rm shell}^{\rm (pair)}$ is even more dramatic. Therefore,
 the contribution from
quasibound states to the pairing correction is a source of another
uncontrolled correction to the total energy in the
macroscopic-microscopic method.

The total energy of   the shell-correction method, Eq.~(\ref{Eshell}),
is divided into three parts:
\begin{equation}\label{ourE}
E = E_{\rm macr} + E_{\rm shell}^{\rm (s.p.)} + E_{\rm shell}^{\rm (pair)},
\end{equation}
where $E_{\rm macr}$ is the macroscopic energy of the Yukawa-plus-exponential
model of Ref.~\cite{[Mol81],[Mol81a]},
$E_{\rm shell}^{\rm (s.p.)}$ is the shell correction (\ref{shell}),
and $E_{\rm shell}^{\rm (pair)}$ is the pairing correction
computed using the BCS procedure.
(Results of the LN calculations are fairly similar and will
not be discussed here.)

The predicted one- and two-neutron separation energies for the
neutron-rich tin and lead isotopes
are shown in Figs.~\ref{Sn_S} and \ref{Pb_S}, respectively.
The two-neutron separation energies (top)
were computed directly from binding energies of
even-even nuclei. Since no self-consistent blocking
was performed for odd nuclei, the one-neutron separation energies (bottom)
were approximated
by means of Eq.~(\ref{separ1n}).
As seen in  Figs.~\ref{Sn_S} and \ref{Pb_S},
the influence of the continuum on the calculated
positions of driplines is quite significant. The two-neutron separation
energies
calculated in the WK method are systematically lower than those
obtained in
the SAM and the difference approaches $\sim$1.6~MeV at
$N$=106. This difference results in a shift in the position of
the two-neutron drip line.
According to the semiclassical approach, the nucleus $^{156}$Sn$_{106}$
appears at the two-neutron drip line, while in the SAM calculations
there are several ($\sim$5)
more stable even-even tin isotopes expected. A similar
trend is seen in the position of  the
one-neutron drip line which
in the SAM calculations is overestimated by 2-4 mass units.

\subsection{Fermi-level self-consistency condition}
For the lead isotopes
the Fermi level in the WK method becomes
positive at $N$$\sim$175, and consequently no solution to the
particle number equation (\ref{Npart1}) can be found.
At the same time, the value
of $S_{2n}$ for $N$$\sim$174 is still positive (Fig.~\ref{Pb_S})
and equals to about 1.7 MeV.
This constitutes a
contradiction because, according to Eq.~(\ref{separ2n}),
the values of $S_{2n}$ and
$\lambda_n$ should vanish simultaneously at the two-neutron drip line.

In the self-consistent theory (e.g., HF+BCS or HFB),
the Fermi energy is equal to the
derivative of the total energy
(ground state energy of the even-even system)
with respect to the particle number.
However,
in the shell-correction method this relation is violated due to the
particle-number inconsistency inherent to the macroscopic-microscopic model.
Differentiating both sides of Eq.~(\ref{ourE}) with respect to $N$
(assuming the parameters of the one-body potential to be fixed),
one obtains five different Fermi energies, namely,
\begin{equation}\label{ourL}
\lambda_{\rm tot} = \lambda_{\rm macr} - \tilde\lambda +
\lambda_{\rm s.p.}+\delta\lambda,
\end{equation}
where $\lambda_{\rm tot}$ is related to the neutron separation energy,
$\lambda_{\rm macr}$ represents the macroscopic Fermi energy,
$\tilde\lambda$ (or $\lambda_{\rm sc}$ in the WK method)
is the smoothed  neutron Fermi
energy, $\lambda_{\rm s.p.}$ is the neutron Fermi energy
of the single-particle model obtained from the BCS equations,
and $\delta\lambda$ contains contributions from the smoothed pairing energy
and the proton shell correction. (The quantity $\delta\lambda$ is
small and consequently
is ignored in the following discussion
to simplify the presentation.)
It should be noted here that
the finite-difference approximation to
derivatives
of the single-particle energy
with respect to particle number
(i.e., $\tilde\lambda$, $\lambda_{\rm sc}$, and
$\lambda_{\rm s.p.}$) cannot be applied.
(The deviation between derivatives and finite differences
can be as large as 10\,MeV!). The reason is that in the model
based on the average one-body potential, the variation in the particle number
(say from $N$ to $N$+2) leads to  (i) the variation in the chemical
potential, and (ii) the variation in the average potential itself.
Indeed, the parameters of the average potential depend smoothly
on $Z$ and $N$, and this causes a quite sizeable change in single-particle
level density with particle number.

In self-consistent
approaches based on the two-body Hamiltonian, the requirement
$\lambda_{\rm tot} = \lambda_{\rm s.p.}$
is fulfilled automatically.
In the shell-correction method
this requirement can be referred to as the {\em Fermi-level self-consistency
condition}:
\begin{equation}\label{Lscc}
\lambda_{\rm macr}(Z,N) = \tilde\lambda(Z,N)
{}~~~{\rm or}~~~~ \lambda_{\rm macr}(Z,N) = \lambda_{\rm sc}(Z,N)
\end{equation}
for the SAM or WK methods, respectively.

The parameters
of the microscopic model are usually adjusted to selected
single-particle properties of nuclei close to the beta-stability line, and
the parameters of the macroscopic model are found by a global fit
to masses and
fission barriers. Therefore, it is not surprising that when extrapolating
far from stability, the particle-number dependences
of $\lambda_{\rm macr}$ and $\tilde\lambda$ are different.
In Figs.~\ref{Sn_lam} and \ref{Pb_lam}
the Fermi energies
defined in Eq.~(\ref{ourL}) are shown relative to $\lambda_{\rm macr}$
for the tin and lead isotopes, respectively.
The degree of inconsistency in
$\lambda$ is measured by the magnitude of deviation
$\Delta\lambda \equiv \tilde\lambda-\lambda_{\rm macr}$.
It is seen that $|\Delta\lambda|$ is rather small around $N$=82
and $N$=126 for the tin and lead isotopes, respectively, and
it reaches the value of about $\Delta\lambda$=1\,MeV at
drip lines.

The condition (\ref{Lscc}) defines a certain coupling between the
macroscopic and microscopic parameters  in
the shell-correction method.
Since $\Delta\lambda$ varies very smoothly as a function of $N$
(see Figs.~\ref{Sn_lam} and \ref{Pb_lam}),
the condition (\ref{Lscc})
it equivalent to
\begin{equation}\label{conditions}
\lambda_{\rm macr}(Z,N_{\rm min}) = \lambda_{\rm sc}(Z,N_{\rm min}),
{}~~\lambda_{\rm macr}(Z,N_{\rm max})=0,
\end{equation}
where, for a given atomic number $Z$,  $N_{\rm min}$
and $N_{\rm max}$ are neutron numbers corresponding to
 proton- and neutron-drip-line isotopes,
respectively. ($N_{\rm max}$ can be computed from Eq.~(\ref{Npart1})
after putting $\lambda_{\rm sc}=0.$)

The consistency between the microscopic and macroscopic
part of the energy in the shell-correction method has been discussed by
several authors. In particular,
Myers noticed \cite{[Mye70]}, on the basis of the droplet model, that
the parameters of the single-particle model (such as radius, potential
depth, diffuseness) should be related to the parameters of the macroscopic
model.
 In the global mass calculations by M\"oller {\it et al.}
\cite{[Mol93]} the parameters defining the single-particle Hamiltonian
do depend on the droplet model constants. In particular,
as in Eq.~(\ref{potvnp}), the depth
of the folded-Yukawa potential depends linearly on the average
bulk nuclear asymmetry of the droplet model, $\bar\delta$.

To illustrate condition (\ref{conditions}),
we performed
calculations based on the WS average potential with parameters
of ref.\,\cite{[Bol72]}, adjusted according
to Myers\,\cite{[Mye70]}.
Namely, we used:
 \begin{equation}
    a=a_{\rm SO}=0.9 \frac{\ln 5}{\ln 9} \;\mbox{\rm [fm]}
 \end{equation}
for the diffuseness, and
 \begin{equation}
    r_\circ = r_{\circ-{\rm SO}} = R_0\left[1-(a/R_0)^2\right]/A^{1/3}
 \end{equation}
with
 \begin{equation}
    R_0=R_{\rho}+0.82-0.56/R_{\rho}
 \end{equation}
for the radius. The radius $R_{\rho}$ and the
depth of the potential
 \begin{eqnarray}
    R_{\rho} &=& 1.16 A^{1/3} (1+\bar{\epsilon}) \;
				       \mbox{\rm [fm]},  \\
    V_{n,p}  &=& (52.5\mp 48.7 \bar{\delta}) \;\mbox{\rm [MeV]}\label{potvnp}
 \end{eqnarray}
depend on the relative neutron excess
coefficient $I$ through
 \begin{eqnarray}
    \bar{\delta}    &=& \frac{I(1-0.0056 A^{1/3})+0.0028 A^{1/3}
		        (1+I^2)}{1+3.15/A^{1/3}}, \nonumber         \\
    \bar{\epsilon}  &=& -\frac{0.147}{A^{1/3}}+0.33 {\bar{\delta}}^2
			+0.00062 A^{2/3} (1-I)^2.
 \end{eqnarray}
Figure \ref{bolsterli} shows
$\Delta\lambda$=$\lambda_{\rm sc}-\lambda_{\rm macr}$
for the lead isotopes. Different curves were obtained
by multiplying
by a factor $b$
(=1.00,\,1.05,\,1.10,\,1.15,\,1.20)
the neutron excess coefficient $I$
entering through $\bar\delta$
expression
(\ref{potvnp}). (Direct renormalization of $\bar\delta$
is much less convenient since it leads to the modification of
the isoscalar part of the potential depth.)
 As can be seen from Fig.~\ref{bolsterli},
the standard value, $b$=1, gives the values of
$\lambda_{\rm sc}$ too small, in particular
when the  neutron number approaches $N_{\rm max}$; the resulting value,
$N_{\rm max}$=194, is much larger than that of the
macroscopic model, $N$=182
(as indicated in Fig.\,\ref{bolsterli}).
However, by increasing
the value of $b$ by 10\%
conditions (\ref{conditions}) can be met
for the considered lead isotopes.
It remains to be investigated whether the Fermi-level
 self-consistency condition can be
consistently fulfilled for other isotopes.

\section{Conclusions}

The main objective of this study was to investigate the influence of
the particle continuum on the shell-correction energies of the
macroscopic-microscopic method. The shell corrections
obtained by means of the standard averaging method were compared with
those calculated with the semiclassical
Wigner-Kirkwood expansion technique. The systematic error
in $E_{\rm shell}$,
due to the particle continuum,
can be as large as several MeV at the neutron drip line.
According to our calculations, this error depends weakly on deformation.
This suggests a possibility of renormalizing $E_{\rm shell}$ only at the
spherical shape.

As demonstrated
in our study, the use of quasibound states as physical resonances
can lead to serious deviations when extrapolating off beta stability.
In particular, the associated theoretical error in  predicted particle
drip lines can be as large as $\Delta N$=10.
We also emphasized the role of the
self-consistency condition between the microscopic and macroscopic
Fermi energies. If this condition is violated,
the relation between the Fermi
energy and the separation energy is lost.

We hope that our work will be helpful for future global calculations
of nuclear masses in the framework of the one-body
(macroscopic-microscopic)
description.  As will be discussed in the
forthcoming study
based on self-consistent two-body procedures
\cite{[Dob94a]}, there are also other uncertainties
related,
e.g., to the choice of the effective interaction. A deep
understanding of the coupling between single-particle and pairing channels,
and between discrete states and particle continuum,
is a key to the physics of drip-line systems and a serious challenge
for future work.

Oak Ridge National
Laboratory is managed for the U.S. Department of Energy by Martin
 Marietta Energy Systems, Inc. under contract No.
DE--AC05--84OR21400.
The Joint Institute for Heavy Ion
 Research has as member institutions the University of Tennessee,
Vanderbilt University, and the Oak Ridge National Laboratory; it
is supported by the members and by the Department of Energy
through Contract No. DE--FG05--87ER40361 with the University
of Tennessee.  Theoretical nuclear physics research at the
University of Tennessee is supported by the U.S. Department of
Energy through Contract No. DE--FG05--93ER40770.
This research was also supported by the Polish Committee
for Scientific Research under Contract No. 20450~91~01.

\appendix

\section{Semiclassical Approximation to the Woods-Saxon Model}\label{Ap1}

The semiclassical equations (\ref{Npart1})
and (\ref{part1}) are defined through the high-order derivatives of  average
potentials entering the WS Hamiltonian.
In order to calculate $\bbox{\nabla}V$ and $\nabla^2V$, we took
advantage of certain geometric relations
specific to the definition of the WS potential (\ref{WS}).
Figure~\ref{dist} shows the surface geometry typical for an  axial
system; here the nuclear surface,  $r$=$R(\theta)$,
is given by Eq.~(\ref{Ro}).
Given the radius-point $\bbox{r}$=$\bbox{r}_{OB}$ (see Fig.~\ref{dist})
the function $\xi(\bbox{r})$
represents the distance $\pm|\bbox{r}_{AB}|$ [taken with the minus (plus)
sign inside (outside) the surface $R(\theta)$]. Denoting by
$\bbox{n}$ [$\bbox{n}$=$\bbox{\nabla}\xi(\bbox{r})$]  and
$\bbox{t}$ the normal and tangent  unit vectors to the surface
at point $A$, respectively, one can write
\begin{equation}\label{basic}
\bbox{r} = \bbox{R} + \bbox{n}\xi(\bbox{r}),
\end{equation}
where $\bbox{R}$=$\bbox{r}_{OA}$.
By acting with the  gradient operator on  both sides of Eq.~(\ref{basic}),
one obtains
\begin{equation}\label{basic1}
(\bbox{\nabla n})\xi(\bbox{r}) = 2 - \bbox{\nabla R}.
\end{equation}
In order to calculate $\bbox{\nabla R}$, one computes the gradient
in the Frenet frame defined by three unit vectors
$\bbox{n}$, $\bbox{t}$, and
$\bbox{b}$=$\bbox{t}\times\bbox{n}$. The result is
\begin{equation}
\bbox{\nabla R} = \frac{\varrho}{\varrho+\xi} + \frac{\eta}{\eta+\xi},
\end{equation}
where the curvature radius, $\varrho$, and $\eta$=$|\bbox{r}_{AC}|$ are
given by
\begin{equation}\label{basic2}
\varrho=\frac{\left[R^2+{R'}^2\right]^{3\over 2}}
{R^2 + 2{R'}^2 - R\,R''},
{}~~\eta=\frac{\left[R^2+{R'}^2\right]^{1\over 2}}
{1 - R'\cot\theta/R},
\end{equation}
and $R'$=$dR/d\theta$ and
$R''$=$d^2R/d\theta^2$.
Finally, by means of Eqs.~(\ref{basic1})-(\ref{basic2}), one obtains
\begin{equation}\label{basic3}
\bbox{\nabla n} = \nabla^2\xi =
\frac{1}{\varrho+\xi} + \frac{1}{\eta+\xi}.
\end{equation}
In the spherical case, $\xi(r)=r-R$, $\varrho=\eta=R$, and
$\bbox{\nabla n}=2/r$, as expected.
For  given $\bbox{r}$,
the distance $\xi(\bbox{r})$ and the vector $R$ are calculated
numerically by solving the equation $d|r-R({\theta})|/d\theta$=0.
By denoting
\begin{equation}
V^{(n)}(\bbox{r}) \equiv \frac{d^n V}{d\,\xi^n},
\end{equation}
one can compute
the derivatives appearing in Eqs.~(\ref{Npart1}) and (\ref{part1}).
For example:
\begin{eqnarray}\label{deriv}
\bbox{\nabla}V^{(n)} & = & V^{(n+1)}\bbox{n},\nonumber \\
(\bbox{\nabla}V^{(n)})^2 & = & (V^{(n+1)})^2,\nonumber\\
\nabla^2 V^{(n)} & = & V^{(n+2)} + V^{(n+1)}\nabla^2 \xi.
\end{eqnarray}
The derivatives higher than 2 involve quantities such
as $\bbox{n}\cdot\bbox{\nabla}(\nabla^2\xi)$ or $\nabla^4 \xi$ which
are computed numerically by means of the finite difference method.
It is known that the standard parameterizations of the nuclear shape
give rise to a singularity of higher derivatives at $r$=0, or
to their large variations at $\theta$=0$^\circ$ or
$\theta$=90$^\circ$ \cite{[Dam69],[Got71]}.
In order to achieve better accuracy of higher-order terms
of the WK expansion, it has ben assumed that the
WS potential is constant for the values of $\xi(\bbox{r})$$<$$\xi_{crit} =
-7a$.
We have checked that
up to $\beta_2$$\sim$0.7
the results are stable with respect to variations of
$\xi_{crit}$.

It is interesting to note that although $(\bbox{\nabla}V)^2$=$(V^{(1)})^2$
is a simple analytic function of $\xi(\bbox{r})$, the Laplacian
$\nabla^2 V$ depends both on $\xi(\bbox{r})$ and
$\theta$, see Eqs.~(\ref{deriv})
and (\ref{basic3}). This will lead to
slight modifications of
some discussion in Refs.~\cite{[Dud84a],[Ben89a]}, where
$\nabla^2 V$ was assumed to depend solely on the distance $\xi(\bbox{r})$.

Special care should be taken when integrating singularities
at $\lambda_{\rm sc}=U$
in Eqs.~(\ref{Npart1}) and (\ref{part1}).
A practical way of handling singularities at
at $\bbox{r}_{\rm sc}$
is to employ the identity:
\begin{eqnarray}\label{Npart6}
\int_{\bbox{r}_{\text{min}}}^{\bbox{r}_{\rm sc}} &d^3r&
\frac{\Phi(\bbox{r})}{\sqrt{\lambda_{\rm sc}-U}} =
\int_{\bbox{r}_{\text{min}}}^{\bbox{r}_{\rm sc}} d^3r
\frac{\Phi(r,\Omega)-\Phi(r_{\rm sc}(\Omega),\Omega)}{\sqrt{\lambda_{\rm
sc}-U}}
\nonumber \\
 & + &
2\int d\Omega\, r^2_{\text{min}}(\Omega)
\frac{\Phi(r_{\rm sc}(\Omega),\Omega)}
{\left({\partial U}/{\partial
r}\right)_{|r_{\text{min}}(\Omega)}}
\sqrt{\lambda_{\rm sc}-U(r_{\text{min}}(\Omega),\Omega)}
\nonumber\\
 & + &
2\int d\Omega\,\Phi(r_{\rm sc}(\Omega),\Omega)
\int_{r_{\text{min}}(\Omega)}^{r_{\rm sc}(\Omega)} dr\,
\frac
{r\sqrt{\lambda_{\rm sc}-U(r,\Omega)}}
{\left({\partial U}/{\partial
r}\right)}
\left[2-\frac{r\left({\partial^2 U}/{\partial
r^2}\right)}{\left({\partial U}/{\partial
r}\right)}
\right],
\end{eqnarray}
where $\Phi(\bbox{r})$ is a function of $\bbox{r}=(r, \Omega)$ and
the integration is performed over the volume surrounded
by the surfaces $r=r_{\text{min}}(\Omega)$ and
$r=r_{\rm sc}(\Omega)$.
In the above equation, the  singularity at
$\bbox{r}_{\rm sc}$ has been removed at the expense of generating
singularity inside the classical region
at $(\bbox{\nabla}U)^2$=0. (Such a
situation happens for the protons, due to different radial behaviors
of WS and Coulomb potentials.)
In practice, however, the identity (\ref{Npart6}) is only used in
the narrow region around $\bbox{r}_{\rm sc}$ where $(\bbox{\nabla}U)^2$
is never zero.

\section{Coulomb Potential  of the Fermi Distribution}\label{Coull}

The Coulomb potential generated by the WS (Fermi)
charge  distribution (\ref{formfactor})
is given by
\begin{equation}\label{Coulomb}
V_{\text{C}}(\bbox{r}) = \int d^3r'
\frac{\rho_\circ}
{1+{\exp}\left[\xi(\bbox{r'})/a\right]}\frac{1}{|\bbox{r}-\bbox{r}'|}.
\end{equation}
In the axial case this integral can be reduced
to two dimensions (with integrand involving  complete
elliptic integrals).
Since
\begin{equation}
\nabla^2 V_{\text{C}}(\bbox{r}) = - 4\pi \rho_{\text{C}}(\bbox{r}),
\end{equation}
the contributions from the Coulomb potential to higher-order
derivatives
in Eqs.~(\ref{Npart1})-(\ref{part1}) are calculated easily
using expressions derived in Appendix~\ref{Ap1}.

For a spherical shape, the angular integrations in Eq.~(\ref{Coulomb})
can be performed explicitly to give
\begin{equation}
V_{\text{C}}(r) = \frac{4\pi}{r}\int_{0}^r dr'\,r'^2\rho_{\text{C}}(r')
+ {4\pi}\int_{r}^{\infty} dr'\,r'\rho_{\text{C}}(r').
\end{equation}
Finally, the radial integrals are computed
by means of identity
\begin{equation}
\frac{1}{1+\exp[(r-R)/a]} = \left\{
\begin{array}{ll}
1+\sum_{n=1}^{\infty}(-1)^n\exp[-n(R-r)/a] & \mbox{if $r<R$,} \\
-\sum_{n=1}^{\infty}(-1)^n\exp[-n(r-R)/a] & \mbox{if $r>R$}.
\end{array}
\right.
\end{equation}
The resulting fast-converging series gives $V_{\text{C}}$ with desired
accuracy.


\begin{figure}
\caption{Schematic illustration of coupling between bound states
and particle
continuum  in drip-line nuclei.
The potential  wells are represented by the average Woods-Saxon field
(plus Coulomb potential for protons).
\label{wells}}
\end{figure}

\begin{figure}
\caption{Single-particle energies
of the WS Hamiltonian
(\protect\ref{WSH})
for neutrons (top) and protons (bottom)
in $^{120}$Sn
 as functions of $N_{\rm osc}$ (the number of
harmonic-oscillator quanta included in the basis).
The solid (dashed) lines correspond to $\pi=+$ ($\pi=-$) orbitals.
The left portion
displays the results for spherical shape
($\beta_2$=0); here every single-particle orbital is $(2j+1)$-fold
degenerate. The results characteristic of deformed shapes
($\beta_2$=0.4) are shown in the right portion; each orbital
is 2-fold degenerate due to time-reversal symmetry.
\label{Fig_Sn}}
\end{figure}

\begin{figure}
\caption{Neutron shell correction for $^{100}$Zn (top)
and $^{122}$Zr (bottom)  for three quadrupole deformations,
$\beta_2$=0, 0.3, 0.6,
as a function of the number of harmonic-oscillator
quanta included in the basis.
 For the SAM calculations we have used
$\gamma$=1.2\,$\hbar\omega_\circ$=49.2\,MeV/$A^{1/3}$
and an eighth order curvature correction ($p$=8). All single-particle levels
lying below a cut-off energy of
$\epsilon_{cut}$=3.2\,$\hbar\omega_\circ$=131.2\,MeV/$A^{1/3}$
above the Fermi level were
included.
\label{Fig_ZrN}}
\end{figure}

\begin{figure}
\caption{Neutron average level density $\tilde{g}$ for $^{100}$Zn (top)
and $^{122}$Zr (bottom)  for $\beta_2$=0.3
as a function of single-particle energy $\epsilon$.
 For the SAM calculations we have used
the value of $\gamma$=1.2\,$\hbar\omega_\circ$
and
$\epsilon_{cut}$=3.2\,$\hbar\omega_\circ$. The average
Fermi energies $\tilde\lambda$  are indicated
by stars.
\label{gavr}}
\end{figure}

\begin{figure}
\caption{Total shell correction
at $\beta_2$=0
as a function of the smoothing range
$\gamma$ (in units of $\hbar\omega_\circ$)
for  $^{100}$Zn (top)
and $^{122}$Zr (bottom). The three different curves correspond to
$p$=6, 10, and 14. The number of basis states used in the diagonalization
is $N_{\rm osc}$=12.
\label{plateaup}}
\end{figure}

\begin{figure}
\caption{Neutron shell corrections
of the WS model,
${E}_{\rm shell}$(SAM) and ${E}_{\rm shell}$(WK), given
by Eqs.~(\protect\ref{Eplateau}) and (\protect\ref{part1}), respectively,
as a function of quadrupole deformation for
$^{100,120,124}$Zr (top) and $^{166,186,206,226}$Os (bottom).
The SAM smoothing was performed with
$N_{\rm osc}$=12, $\gamma$=$\hbar\omega_\circ$, and $p$=8.
\label{WKS}}
\end{figure}

\begin{figure}
\caption[FF]{Neutron BCS pairing gaps for the even tin isotopes as a function
of $N$ computed
using the average pairing strength $G_{\rm avg}$ \cite{[Mol92b]}
(solid line) or
the fitted pairing strength $G_{\rm fit}$, (Eq.~(\protect\ref{Gfit})
for $G_\circ$=19.2\,MeV and
$G_1$=$-$7.4\,MeV, dashed line).
The single-particle spectrum
(including quasibound states) was obtained by diagonalizing the WS
Hamiltonian in $N_{\rm osc}$=12 (top) and
$N_{\rm osc}$=40 (bottom) oscillator shells.
\label{DELTAG}}
\end{figure}

\begin{figure}
\caption{One- and two-neutron separation energies for the
even
neutron-rich tin isotopes predicted in the shell-correction method
with the WS average potential and the Yukawa\--plus\--exponential
macroscopic model. The results based on the semiclassical
Wigner-Kirkwood method
(solid line) are compared with those obtained using the standard
averaging including quasibound states. The  smoothing was performed with
$N_{\rm osc}$=12 (dotted line),
or $N_{\rm osc}$=40 (dash-dotted line),
$\gamma$=$\hbar\omega_\circ$, and $p$=8.
\label{Sn_S}}
\end{figure}

\begin{figure}
\caption{Same as in Fig.~\protect\ref{Sn_S} except for
the neutron-rich lead isotopes.
\label{Pb_S}}
\end{figure}

\begin{figure}
\caption{Fermi energies of
Eq.~(\protect\ref{ourL}) relative to $\lambda_{\rm macr}$
as a function of $N$
for
the neutron-rich tin isotopes.
The degree of the inconsistency in
$\lambda$ is measured by the magnitude of deviation
$\tilde\lambda-\lambda_{\rm macr}$ (SAM variant)
or
$\lambda_{\rm sc}-\lambda_{\rm macr}$ (WK variant).
The particle numbers
at which   $\lambda_{\rm tot}$=0, $\tilde\lambda$=0,
$\lambda_{\rm sc}$=0,
$\lambda_{\rm macr}$=0, and $\lambda_{\rm s.p.}$=0 are indicated by stars.
\label{Sn_lam}}
\end{figure}

\begin{figure}
\caption{Same as in Fig.~\protect\ref{Pb_lam} except for
the neutron-rich lead isotopes.
\label{Pb_lam}}
\end{figure}

\begin{figure}
\caption{Semiclassical Fermi energy,
$\lambda_{\rm sc}$,
relative to
$\lambda_{\rm macr}$
for the lead isotopes.
The parameters of the WS model were taken according to
ref.~\protect\cite{[Bol72]}.
The curves correspond to different
values of parameter $b$ multiplying the neutron excess coefficient $I$
entering through
$\bar{\delta}$
the formula
(\protect\ref{potvnp}) for the neutron potential depth.
The particle numbers $N$=$N_{\rm max}$ at which
$\lambda_{\rm sc}$=0 and
$\lambda_{\rm macr}$=0 [Eq.~(\protect\ref{conditions})]
 are indicated by stars.
\label{bolsterli}}
\end{figure}

\begin{figure}
\caption{Geometric relations between a vector-point $\protect\bbox{r}$, the
distance function $\xi(\protect\bbox{r})$, and the nuclear surface,
$R(\theta)$. The unit vectors $\protect\bbox{n}$,
$\protect\bbox{t}$, and
$\protect\bbox{b}$ (normal,
tangent, and binormal vector fields of $R$) represent
the Frenet frame field on $R$.
\label{dist}}
\end{figure}

\end{document}